\documentclass{PoS}

\def \be{\begin{equation}}
\def \ee{\end{equation}}
\def \bea{\begin{eqnarray}}
\def \eea{\end{eqnarray}}
\def \ben{\begin{enumerate}}
\def \een{\end{enumerate}}
\def \Bbar{\overline{\kern -0.24em B}}
\usepackage{epsfig}

\usepackage{multicol}
\usepackage{rotating}

\title{Correlations between high-$p_T$ and flavour physics\footnote{CERN-PH-TH/2009-206, MZ-TH/09-43 } }

\ShortTitle{Correlations between high-$p_T$ and flavour physics}

\author{\speaker{Tobias Hurth}\\
       \\ CERN, Dept. of Physics, Theory  Division, CH-1211 Geneva 23, Switzerland\\
       Institute for Physics, Johannes Gutenberg-University, D-55099 Mainz, Germany\\         
 E-mail: \email{tobias.hurth@cern.ch}}

\author{Werner Porod\\
	\\ Inst.\ f.\ Theoretische Physik und Astrophysik,
       Uni.\ W\"urzburg, D-97074 W\"urzburg, Germany \\
	E-mail: \email{porod@physik.uni-wuerzburg.de}}

\abstract{ 
Squark and gluino decays are governed by the same
mixing matrices as the contributions to flavour violating loop
transitions of $B$-mesons.  This allows for possible direct correlations between
flavour non-diagonal observables in $B$ and high-$p_T$ physics.
 The present bounds on squark mixing, induced by the low-energy data 
 on $b \to s$ transitions, still allow for large contributions to flavour violating 
 squark decays at tree level.  Due  to the restrictions in flavour tagging at the LHC, additional
information from future flavour experiments will be necessary to
interpret those LHC data properly.  Also the measurement of
correlations between various squark decay modes at a future ILC would
provide information about the flavour violating parameters.
}

\FullConference{European Physical Society Europhysics Conference on High Energy Physics,
EPS-HEP 2009,\\
		 July 16 - 22 2009\\
		 Krakow, Poland}

\begin{document}

\section{Introduction}

Rare $B$ and kaon decays (for a review see
\cite{Hurth:2007xa,Hurth:2003vb}) representing loop-induced processes
are highly sensitive probes for new degrees of freedom beyond the SM
establishing an alternative way to search for new physics. 
The day the existence of new degrees of
freedom is established by the Large Hadron Collider (LHC), the
 present stringent flavour bounds will translate in first-rate
information on the new-physics model at hand.

Thus, within the next decade an important interplay of flavour and
high-$p_T$  physics most probably will take place. For example, within
supersymmetric extensions of the SM, the measurement of the flavour
structure is directly linked to the crucial question of the
supersymmetry-breaking mechanism as the soft SUSY breaking terms are
the source of flavour structures beyond the SM.  LHC has the potential 
to  discover
strongly interacting supersymmetric particles up to a scale of 2 TeV
and to  measure several of their properties 
\cite{Ball:2007zza,delAguila:2008iz,Buchalla:2008jp,Aad:2009wy}. 
This information can be used for a refined analysis of flavour physics
observables indicating possible flavour structures and, thus, give
important information for distinguishing between models of
supersymmetry breaking.

Data from $K$ and $B_d$ physics show  that new sources of flavour violation
in \mbox{$s \rightarrow d$} and \mbox{$b \rightarrow d$} 
are strongly constrained, while 
the possibility of sizable  new 
contributions to \mbox{$b \rightarrow s$}  
remains open. 
In  \cite{Hurth:2003th,Hurthnew}
we analysed  flavour violating squark and gluino decays and showed  that they
can still be typically  of order 10\% despite the stringent
constraints from low energy data. For a related study see also
\cite{Bartl:2009au}.

\section{Phenomenological analysis}

Within the Minimal Supersymmetric Standard Model (MSSM) there are two
new sources of flavour changing neutral currents (FCNC), namely new
contributions which are induced through the quark mixing as  in the
SM and generic supersymmetric contributions through the squark
mixing.
In the latter case, flavour violation is induced by off-diagonal elements of
the squark mass matrices. We normalize them by   
the average of the diagonal elements (trace of the
mass matrix divided by six) in the up and down sector, denoted by
$m_{\tilde{q}}^2$, to be independent of the SUSY point under study.
The observables can then be studied as a function
of the normalized off-diagonal elements $(i \ne j)$: 
\begin{equation} 
\delta_{LL,ij} = \frac{(M^2_{\,f,\,LL})_{ij}}{m^2_{\tilde{q}}}\,, 
\delta_{f,RR,ij} = \frac{(M^2_{\,f,\,RR})_{ij}}{m^2_{\tilde{f}}}\,, 
\delta_{f,LR,ij} = \frac{(M^2_{\,f,\,LR})_{ij}}{m^2_{\tilde{f}}}\,,
\delta_{f,RL,ij} = \frac{(M^2_{\,f,\,RL})^\dagger_{ij}}{m^2_{\tilde{f}}}\,.
\label{deltadefb}
\end{equation}
where $f$ is either $u$ or $d$ for $u$-squarks and $d$-squarks, respectively.
A consistent analysis of the bounds  should 
also include interference effects between the various contributions, 
namely the interplay between the various sources of flavour violation 
and the interference effects  of SM and various new-physics contributions~\cite{Besmer:2001cj}.

We first fix the flavour-diagonal set of
parameters and then we vary the flavour-nondiagonal parameters and
explore the bounds on those parameters by theoretical and experimental
constraints.  For flavour-diagonal parameter we use 
the popular SUSY benchmark point SPS1a'
\cite{Aguilar-Saavedra:2005pw}, for a comparison with other
study points see \cite{Hurthnew}. 
 SPS1a' contains the lightest spectrum with squarks around
500 GeV and $m_{\tilde g}$ around 600 GeV and $\tan\beta=10$, being
consistent with WMAP data \cite{Spergel:2006hy} and measurements of
the anomalous magnetic moment of the muon.

On the flavour-nondiagonal parameter set we pose the theoretical 
vacuum stability bounds, all constraints  from electroweak precision data,
the squark Tevatron bounds. Finally we also use 
the explicit experimental constraints from the  most important
 flavour observables, namely  $Br(\bar B \to X_s \gamma)$,
 $BR(\bar B \to  X_s l^+l^-)$, $BR(B_s \to \mu^+ \mu^-)$, and $\Delta M_{B_S}$
 ~\cite{Misiak:2006zs,Lunghi:2006uf,Cho:1996we,Hurth:2003dk,Buras:2002vd,Huber:2007vv}:
Those bounds include experimental {\it and} theoretical errors which
are linearly added.  Explicitly our bounds are the experimental $95\%$
bounds where twice the SM uncertainty is added in order to take into account
uncertainties of the new physics contributions in a conservative
way. We have also checked that the recent experimental data on $B \to
\tau \nu$ do not give additional constraints.
For the numerical evaluation we use an updated version of {\tt SPheno}
\cite{Porod:2003um} which has been extended to accept flavour mixing
entries in the sfermion mass matrices.  

\begin{table}[t] 
\caption{Branching ratios larger than 1\% for  two study points. The flavour
diagonal entries are according to SPS1a'. $\tilde u_i$ decays are like in SPS1a'
\cite{Aguilar-Saavedra:2005pw} and in both scenarios 
BR$(\tilde d_3 \to \tilde \chi^0_1 d) = 99.1$\%.}
\label{tab:BR}
%\begin{center}
\vskip 2mm %\hskip6mm
\begin{tabular}{|c||r|r|r||r|r|r|}
\hline
decaying & \multicolumn{6}{|c|}{final states and corresponding branching ratios in \% for.}  \\
 particle & \multicolumn{3}{|c||}{I. $\delta_{LL,23}=0.01,\delta_{D,RR23}=0.1$}
        & \multicolumn{3}{|c|}{II. $\delta_{LL,23}=0.04,\delta_{D,RR23}=0.45$}
\\ \hline
$\tilde d_1 \to $  
   & $\tilde \chi^0_1 b$\,,\, 4.4 & $\tilde \chi^0_2 b$\,,\, 29.8 &  $\tilde \chi^-_1 t$\,,\, 37.0
   & $\tilde \chi^0_1 s$\,,\, 36.8 & $\tilde \chi^0_1 b$\,,\, 42.2 &   $\tilde \chi^0_2 b$\,,\, 10.9
   \\ 
   & $\tilde u_1 W^-$\,,\,  27.7 & &
   &  $\tilde \chi^-_1 t$\,,\, 9.6 & & \\
\hline
$\tilde d_2 \to $  
   & $\tilde \chi^0_1 s$\,,\, 8.0 &  $\tilde \chi^0_1 b$\,,\, 6.4 &  $\tilde \chi^0_2 b$\,,\, 19.0
   & $\tilde \chi^0_1 b$\,,\, 2.1 & $\tilde \chi^0_2 b$\,,\, 27.3 &  $\tilde \chi^-_1 t$\,,\, 34.6
   \\ 
   & $\tilde \chi^0_3 b$ \,,\, 1.1 & $\tilde \chi^0_4 b$ \,,\, 1.8 & $\tilde \chi^-_1 t$\,,\, 24.6
   &  $\tilde u_1 W^-$\,,\,  33.2 & & \\
   & $\tilde u_1 W^-$\,,\,  38.9 & &   &   & & \\
\hline
$\tilde d_4 \to $  
   & $\tilde \chi^0_1 s$\,,\, 9.1 &  $\tilde \chi^0_1 b$\,,\, 6.3 &  $\tilde \chi^0_2 s$\,,\, 25.3
   & $\tilde \chi^0_1 d$\,,\, 2.3 & $\tilde \chi^0_2 d$\,,\, 31.7 &  $\tilde \chi^-_1 u$\,,\, 59.7
   \\ 
   & $\tilde \chi^-_1 u$\,,\, 2.1 &  $\tilde \chi^-_1 c$\,,\, 47.3 & $\tilde u_1 W^-$\,,\,  4.8 
   &  $\tilde \chi^-_1 c$\,,\, 3.0  & $\tilde \chi^-_2 u$\,,\, 2.3& \\
\hline
$\tilde d_5 \to $  
   & $\tilde \chi^0_1 d$\,,\, 2.3 & $\tilde \chi^0_2 d$\,,\, 31.7 &  $\tilde \chi^-_1 u$\,,\, 59.9
   & $\tilde \chi^0_1 s$\,,\, 2.2 &  $\tilde \chi^0_2 s$\,,\, 30.7 &  $\tilde \chi^-_1 u$\,,\, 2.9
   \\ 
   &  $\tilde \chi^-_1 c$\,,\, 2.8  & $\tilde \chi^-_2 u$\,,\, 2.3& 
  & $\tilde \chi^-_1 c$\,,\, 58.5 &  $\tilde \chi^-_2 c$\,,\, 2.3 &  \\
\hline
$\tilde d_6 \to $  
   & $\tilde \chi^0_1 s$\,,\, 3.1 & $\tilde \chi^0_2 s$\,,\, 30.6 &  $\tilde \chi^-_1 u$\,,\, 2.7
   & $\tilde \chi^0_1 s$\,,\, 19.7 &  $\tilde \chi^0_1 b$\,,\, 18.8 &  $\tilde \chi^0_3 b$\,,\, 2.9
   \\ 
  & $\tilde \chi^-_1 c$\,,\, 58.1 &  $\tilde \chi^-_2 c$\,,\, 2.4 & 
  & $\tilde \chi^0_4 b$\,,\, 2.9  & $\tilde \chi^-_2 t$\,,\, 5.8  & $\tilde g s$\,,\, 2.2  \\
  & & &
  & $\tilde g b$\,,\, 39.8 &$\tilde u_1 W^-$\,,\,  5.5 & \\
\hline
$\tilde g \to$ & $\tilde u_1 t$\,,\, 19.2 & $\tilde u_2 c$\,,\, 8.2 & $\tilde u_3 u$\,,\, 8.3 
               & $\tilde u_1 t$\,,\, 13.5 & $\tilde u_2 c$\,,\, 5.8 & $\tilde u_3 u$\,,\, 5.8 \\
               & $\tilde u_4 u$\,,\, 4.2 & $\tilde u_5 c$\,,\, 4.2 &  
               & $\tilde u_4 c$\,,\, 2.6 & $\tilde u_5 u$\,,\, 2.6 &  \\
          & $\tilde d_1 s$\,,\, 1.4 & $\tilde d_1 b$\,,\, 20.6 &  
          & $\tilde d_1 s$\,,\, 21.1 & $\tilde d_1 b$\,,\, 22.7  &  \\
          & $\tilde d_2 s$\,,\, 6.3 & $\tilde d_2 b$\,,\, 9.0 &  $\tilde d_3 d$\,,\, 8.3
          & $\tilde d_2 b$\,,\, 14.0 & &  $\tilde d_3 d$\,,\, 5.9 \\
          & $\tilde d_4 s$\,,\, 2.3 & $\tilde d_4 b$\,,\, 1.3 &  $\tilde d_6 s$\,,\, 2.8
          & $\tilde d_4 d$\,,\, 2.3 & $\tilde d_5 d$\,,\, 3.3  &  \\
\hline
\end{tabular}
%\end{center}
\end{table}
Now we consider   scenarios with large flavour violating
entries in the squark mass matrices focusing on the mixing between
second and third generation squarks. The crucial point is that those 
entries govern both,  flavour violating low energy observables on the one hand and
squark and gluino decays on the other hand.
To illustrate  the effect of the flavour mixing
parameters on the decay properties of squarks and gluinos,  
we use two study points with squark mixing consistent
with present flavour data and other constraints listed above. 
The two study points chosen are characterized by $\delta_{LL,23}=0.01$
and $\delta_{D,RR23}=0.1$ (point I) and $\delta_{LL,23}=0.04$ and
$\delta_{D,RR23}=0.45$ (point II) respectively. Study point II is
characterized by large cancellations of the SUSY contributions to
$B$-physics observables.  In Table \ref{tab:BR} we give a summary of
the various branching ratios:

The relative size of the branching ratios in Table \ref{tab:BR} can be 
understood by the nature of  the various squarks mass eigenstates.
In point I one finds $\tilde d_1 \simeq \tilde b_L$ with a small
admixture of $\tilde b_R$,  $\tilde d_2 \simeq \tilde b_R$ with small
admixtures of $\tilde s_R$ and $\tilde b_L$, $\tilde d_3 \simeq \tilde
d_R$, $\tilde d_4 \simeq \tilde s_R$ with admixtures of $\tilde s_L$
and $\tilde b_R$, $\tilde d_5 \simeq \tilde d_L$ and $\tilde d_6
\simeq \tilde s_L$ with a small admixture of $\tilde s_R$. Thus,
larger flavour effects are visible in the decays of $\tilde d_2$ and
$\tilde d_4$ where the flavour violating decay branchings ratios
$\tilde d_2 \to \tilde\chi^0_1 s$ and $\tilde d_4 \to \tilde\chi^0_1
b$ are of the order of 10\%. This structure is also the reason for the
relative importance of the flavour violating decays of the gluino.  As
a side remark we note that the flavour violating decays of the first
generation squarks and of the 2nd/3rd generation squarks into the
first generation quarks are due to CKM quark mixing.

In point II the situation is more complicated  due to the larger flavour
mixing parameters.  With respect to the nature of the $d$-type squarks
we find that $\tilde d_1$ and $\tilde d_6$ are strongly mixed states
consisting mainly of $\tilde s_R$ and $\tilde b_R$ with a small
admixture of $\tilde b_L$ whereas the other states are mainly
electroweak eigenstates: $\tilde d_2 \simeq \tilde b_L$, $\tilde d_3
\simeq \tilde d_R$, $\tilde d_4 \simeq \tilde d_L$ and $\tilde d_5
\simeq \tilde s_L$. In this scenario the flavour violating final
states can even reach about  40\% in case of $\tilde d_2 \to
\tilde\chi^0_1 s$ and about 20\% for $\tilde d_6 \to \tilde\chi^0_1
b$. The differences for the gluino decays between these two points is
not only due to the different mixing in the $d$-squark sector but also
due to the different kinematics. 

\section{Impact on LHC}

Large flavour changing decay modes of squarks and gluinos  clearly have  an impact on the discovery
strategy of such particles  as well as on the measurement of the underlying
parameters at the LHC. For example, in mSUGRA points without flavour mixing one
 finds usually that
the left-squarks of the first two generations as well as the right squarks
have similar masses. Large flavour mixing implies that there is a considerable
mass splitting. 
Therefore, the assumption
of almost degenerate masses should be reconsidered if sizable flavour changing
decays are discovered in squark and gluino decays.

\begin{figure}[t]
 \unitlength 1mm
\begin{picture}(17,50)
\put(0,2){\mbox{\epsfig{figure=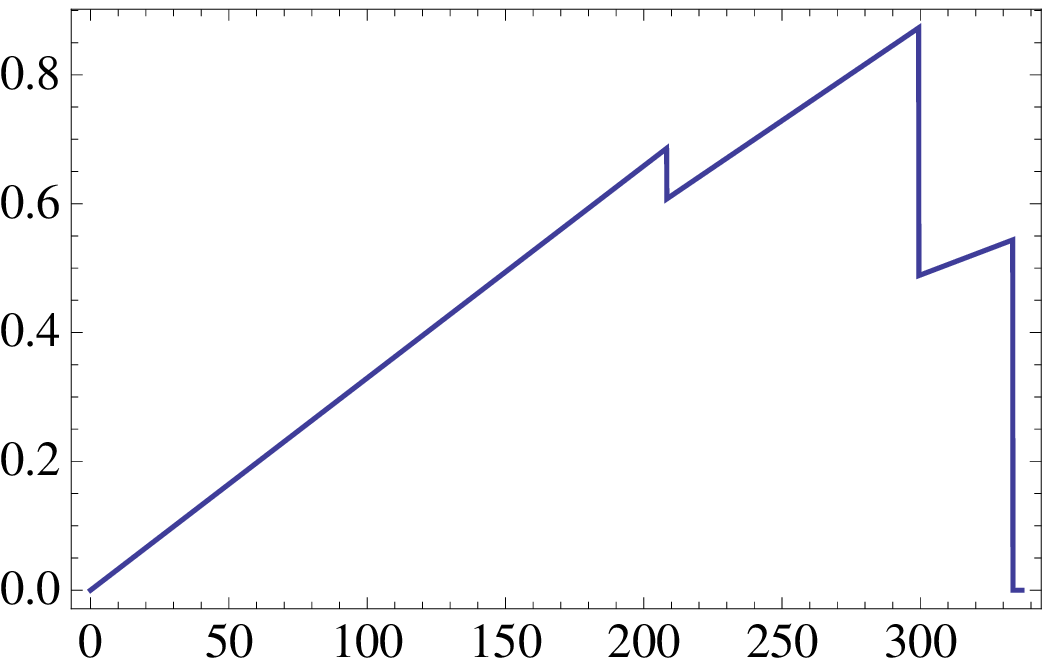,height=4.5cm,width=7cm}}}
\put(37,-2){\mbox{$m_{bb}$}}
\put(0,50){\mbox{{\bf a)} $10^4$ $d$(BR($\tilde g \to b \bar{b} \tilde \chi^0_1)/d m_{bb}$}}
\put(80,2){\mbox{\epsfig{figure=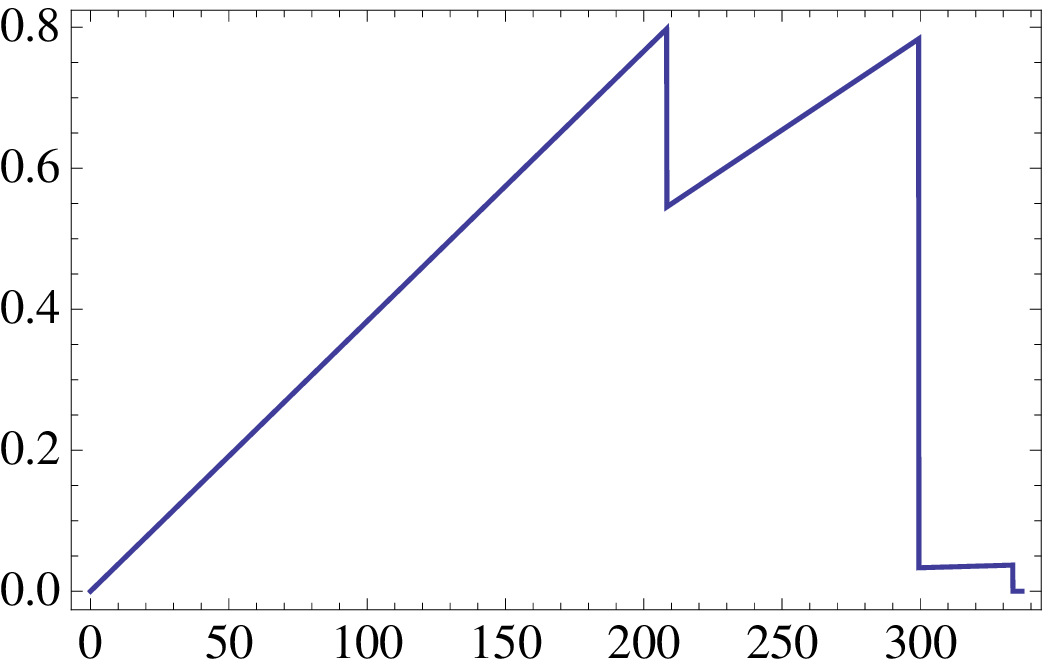,height=4.5cm,width=7cm}}}
\put(127,-2){\mbox{$m_{bs}$}}
\put(80,50){\mbox{{\bf b)} $10^4$ $d$(BR($\tilde g \to b s \tilde \chi^0_1)/d m_{bs}$}}
\end{picture}
\caption{Differential distributions $d$(BR($\tilde g \to b \bar{b} \tilde \chi^0_1)/d m_{bb}$
and $d$(BR($\tilde g \to b \bar{b} \tilde \chi^0_1)/d m_{bb}$
as a function of $m_{bb} = \sqrt{(p_b + p_{\bar{b}})^2}$ ($m_{bs}$) for point 1
defined in table 1. In b) the sum over the charges is shown:
BR($\tilde g \to b \bar{s} \tilde \chi^0_1)$ + BR$(\tilde g \to \bar{b} s \tilde \chi^0_1)$.
 }
\label{fig:bbbar}
\end{figure}

An important part of the decay chains considered for SPS1a' and nearby points
are $\tilde g \to b \tilde b_j \to b \bar{b} \tilde \chi^0_k$ which are
used to determine the gluino mass as well as the sbottom masses or at least
their average value if these masses are close \cite{Branson:2001ak}. 
In the latter analysis the existence of two $b$-jets has been assumed
stemming from this decay chain. In this case the two contributing
sbottoms would lead to two edges in the partial distribution
$d$(BR($\tilde g \to b \bar{b} \tilde \chi^0_1)/d m_{bb}$ where
$m_{bb}$ is the invariant mass of the two bottom quarks.  As can be
seen from Figure \ref{fig:bbbar} there are scenarios where more
squarks can contribute and consequently one finds a richer structure,
e.g.~three edges in the example shown corresponding to study point I.
Such a structure is either a clear sign of flavour violation or the
fact that the particle content of the MSSM needs to be extended.
Moreover, also the differential distribution of the final state $b s
\tilde \chi^0_1$ shows a similar structure where the edges occur at
the same places as in the $b\bar{b}$ spectrum but with different
relative heights. This gives a non-trivial cross-check on the
hypothesis of sizeable flavour mixing. Clearly a detailed Monte Carlo
study will be necessary to see with which precision one can extract
information on these edges. 
Obvious difficulties will be combinatorics
because in general two gluinos or a gluino together with a squark will
be produced and, thus, there will be several jets stemming from light
quarks. However, one could take final states where one gluino decays
into $d$-type squarks and the second into stops or $c$-squarks. In the
second case effective charm tagging would be crucial.

\footnotesize
\section*{Acknowledgement}
This work is supported by the European Network MRTN-CT-2006-035505 'HEPTOOLS'.
W.P.~is  is partially supported by the DFG, project Nr.\ PO 1337/1-1. TH acknowledges
financial support of the ITP at the  University Zurich. 
%T.H. thanks the European Network  MRTN-CT-2006-035505 , HEPTOOLS,  for financial support. 
%W.P.~is  is partially supported by the DFG, project Nr.\ PO 1337/1-1.

%\begin{multicols}{2}

\end{document}